\documentclass[prl,twocolumn,superscriptaddress,showpacs,amsmath,amssymb]{revtex4-1}
\usepackage[dvipdfmx]{graphicx}
\usepackage{mathrsfs}
\usepackage{bm}
\usepackage{siunitx}
\usepackage{dcolumn}
\usepackage{color}
\usepackage[normalem]{ulem}
\def\U#1{{\rm #1}} 
\def\u#1{_{\rm #1}}
\newcommand{\ket}[1]{| #1 \rangle}
\newcommand{\bra}[1]{\langle #1 |}

\newcommand{\expect}[1]{\left\langle #1 \right\rangle} 
\newcommand{\no}[1]{\left\langle: #1 :\right\rangle}

\def\g2{g^{(2)}}

\DeclareRobustCommand{\erase}{\bgroup\markoverwith{\textcolor{red}{\rule[.5ex]{2pt}{0.4pt}}}\ULon}

\begin{document}

\title{Quantum state tomography of qudits via Hong-Ou-Mandel interference}

\author{Yoshiaki Tsujimoto}
\email{tsujimoto@nict.go.jp}
\affiliation{Advanced ICT Research Institute, National Institute of Information and Communications
Technology (NICT), Koganei, Tokyo 184-8795, Japan}
\author{Rikizo Ikuta}
\affiliation{Graduate School of Engineering Science, Osaka University,
  Toyonaka, Osaka 560-8531, Japan}
\affiliation{
  Center for Quantum Information and Quantum Biology, 
    Osaka University, Osaka 560-8531, Japan}
\author{Kentaro Wakui}
\affiliation{Advanced ICT Research Institute, National Institute of Information and Communications
Technology (NICT), Koganei, Tokyo 184-8795, Japan}
\author{Toshiki Kobayashi}
\affiliation{Graduate School of Engineering Science, Osaka University,
  Toyonaka, Osaka 560-8531, Japan}
\affiliation{
  Center for Quantum Information and Quantum Biology, 
    Osaka University, Osaka 560-8531, Japan}
\author{Mikio Fujiwara}
\affiliation{Advanced ICT Research Institute, National Institute of Information and Communications
Technology (NICT), Koganei, Tokyo 184-8795, Japan}

\begin{abstract}
We propose a method to perform the quantum state tomography~(QST) of an $\mathit{n}$-partite qudit state embedded in single photons using the Hong-Ou-Mandel~(HOM) interference between the target state 
and probe state. 
This method requires only passive beam splitters for the HOM interference and removes 
all active optical devices in the target modes to control the measurement bases needed in conventional QST. 
Hence, it is applicable to various degree of freedom of the target 
state without altering the measurement setup.
Moreover, a faithful estimation is realized even with classical probe light such as laser and thermal light.
As a proof-of-principle, we performed the experimental demonstration using a polarization qubit. Regardless of the photon statistics of the probe light, the estimated results of state reconstruction are as accurate as those verified by conventional QST.
\end{abstract}


\maketitle

{\it Introduction.-} The Hong-Ou-Mandel (HOM) interference is a bunching phenomenon based on the bosonic nature of photons~\cite{Hong1987}.
When two photons that are perfectly indistinguishable from each other are mixed using a half beam splitter~(HBS), they always bunch and exhibit the absence of coincidence~\cite{PhysRevLett.100.133601}. This feature has been used in a variety of quantum information processing applications, including quantum computation~\cite{Knill2001,Kaltenbaek2010,Vigliar2021} and 
quantum communication~\cite{Halder2007,Wang2015,Chen2017,RevModPhys.83.33,Yu2020}. 
The decrease in coincidence reflects the degree of indistinguishability, 
i.e., it provides information regarding the probability of overlap of the input photons. Thus, HOM interference enables the estimation of the unknown quantum state of a target photon under test using a probe photon with a known quantum state. Although this concept has been applied for estimating photon spectra~\cite{PhysRevLett.99.123601,Thiel:20}, spatial phase-amplitude structures~\cite{Chrapkiewicz2016}, and photon number distributions~\cite{PhysRevA.79.020101,PhysRevLett.113.070403,Tiedau_2018}, 
there are no reported studies on the reconstruction of the complete density operators of photonic qubits.

\begin{figure}[t]
 \begin{center}
  \includegraphics[width=\columnwidth]{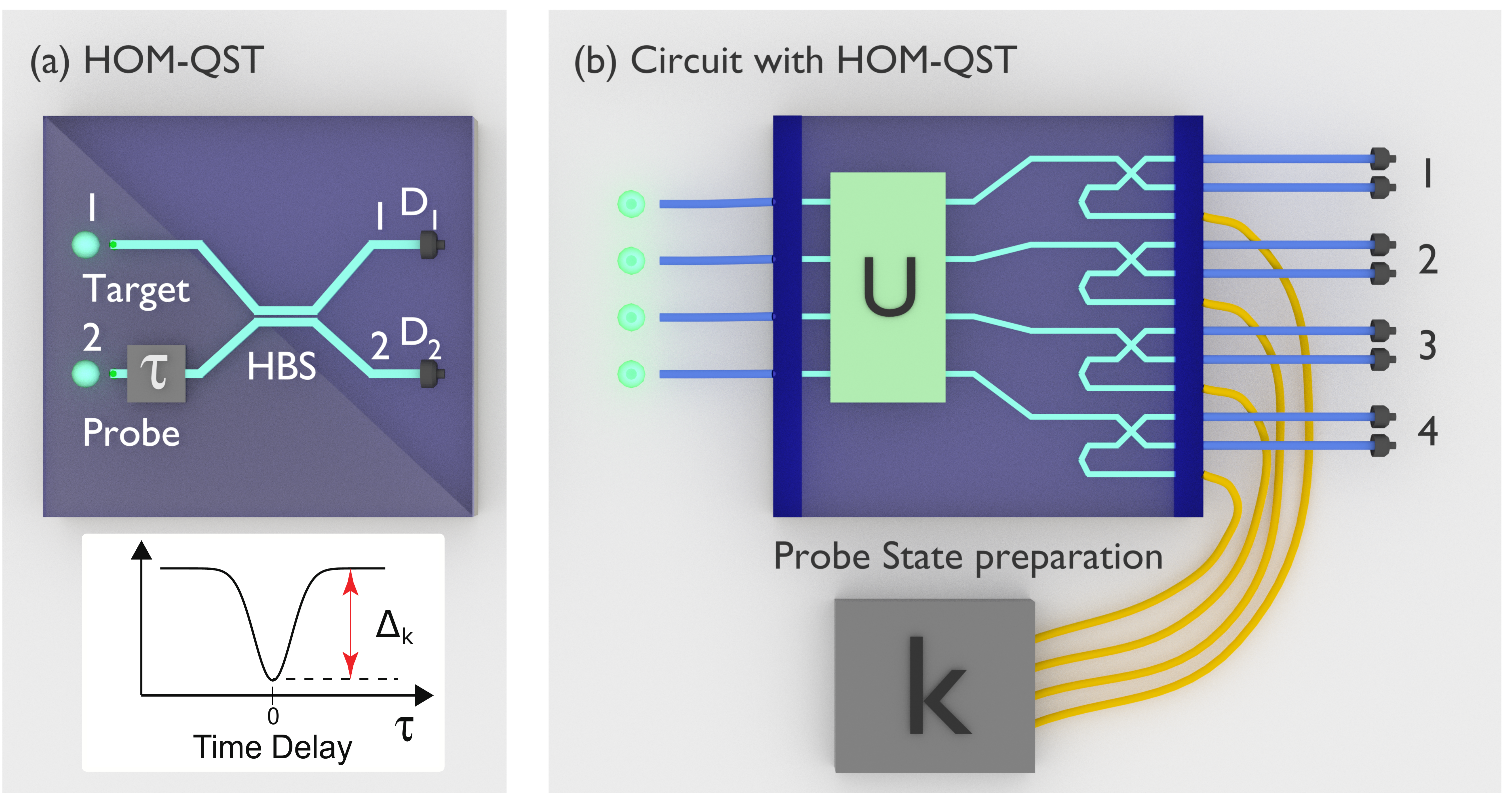}
    \caption{
     (a)~Concept of HOM-QST. 
    (b)~Possible application of HOM-QST. 
        }
 \label{fig:concept}
 \end{center}
\end{figure}

In this Letter, we propose and demonstrate the quantum state tomography (QST) of photonic qubits based on HOM interference, which we refer to as HOM-QST. This concept is illustrated in Fig.~\ref{fig:concept}(a).  
While the conventional QST~\cite{James2001}
is performed by applying a set of projective measurements to the target photon, HOM-QST is performed by interference of the target photon with a set of probe photons.
HOM-QST offers the 
following outstanding advantages over conventional QST: (i)~Only passive optical HBSs are added to the photonic circuits, and there is no need for active control of the target photons under test. This feature enables a low-loss and compact circuit design and is hence ideal for integrated photonic quantum circuits~\cite{Najafi2015,Schuck2016,Gaggero:19,Gyger2021}, as shown in Fig.~\ref{fig:concept}(b). (ii)~This method can be used to estimate the states of various degrees of freedom, such as polarization, time, frequency, and orbital angular momentum, without changing the measurement setup for the target photon. This is because the HBS allows for wide operation over these degrees of freedom~\cite{Harnchaiwat:20,Lingaraju:19,Nagali2009}.
To compensate for the above advantages, HOM-QST needs preparation of the probe light; nonetheless, HOM-QST has the following useful feature:
(iii)~Except for the degree of freedom of interest for QST, there is no need to calibrate the probe photon information, such as the average photon number, photon statistics, and mode mismatch with the target photon. This feature mitigates the measurement cost of the proposed HOM-QST significantly. 

{\it Theory.-} We explain the HOM-QST of a $d$-dimensional qudit
in the Hilbert space spanned by $\{ \ket{m}\}_{m=1,\ldots, d}$
embedded in a single photon. 
The polarization qubit corresponds to the case of $d=2$.  
As shown in Fig.~\ref{fig:concept}(a), we consider the case wherein the target light under test is injected into the HBS from input port~1, and the probe light is injected from input port~2. The typical HOM interferogram~(HOM dip) based on the two-fold coincidence probabilities between detectors $\U{D}_{1}$ and $\U{D}_{2}$ is shown in the inset graph.
The depth of the HOM dip reflects the degree of the bosonic bunching effect due to the indistinguishability of the input photons.
Here, we consider the HOM experiment with the probe light in state $k$ which is a
superposition of the basis states. The depth of the dip is defined by 
$\Delta_k := P_\infty - P_k $, where $P_\infty$ and $P_k$ are the two-fold coincidence probabilities with infinite and zero time delays of the input photons, respectively.

We assume that the detection efficiencies of the detectors are not dependent on $m$ and 
are significantly lower than 1, 
such that the detection probabilities are proportional to the photon numbers 
in the detected modes. 
When we measure the photons without resolving the state $m$, 
the coincidence probability, $P$, between $\U{D}_1$ and $\U{D}_2$ in Fig.~\ref{fig:concept}(a) is given by 
\begin{equation}
P = \eta_{1}\eta_{2}\no{  \sum_m N_{1,m}^\U{out}  \sum_{m'} N_{2,m'}^\U{out} }, 
\label{eq:povm}
\end{equation}
using the normal order representation, where $\hat{N}_{i,m}^\U{out}$ is the number operator of the photon in state $m$ at the output port $i$ of the HBS and 
$\eta_i$ is a detection efficiency of  $\U{D}_i$.

When there is no mode mismatch between the target and probe lights, 
the transformation of the HBS is given by 
$N_{1,m}^\U{out}\rightarrow (a_{1,m}^\dagger + a_{2,m}^\dagger) (a_{1,m} + a_{2,m})/2$ 
and 
$N_{2,m'}^\U{out}\rightarrow (a_{1,m'}^\dagger - a_{2,m'}^\dagger) (a_{1,m'} - a_{2,m'})/2$, 
where $a_{i,m}$ is the annihilation operator of the photon in state $m$ at input port $i$ of the HBS. 
Assuming that 
the target and probe lights have no phase correlations 
and are statistically independent~\cite{Tsujimoto2017,Tsujimoto2018}, $P_k$ is calculated as follows: 
\begin{eqnarray}
P_k 
&=& \frac{\eta_{12}}{4} \left(
\no{ (N_1+N_{2,k})^2 } - 2 \expect{ N_{1,k}}\expect{N_{2,k}} \right) \label{eq:N} \\ 
&=& \frac{\eta_{12}}{4} \left(n_s^2 \g2_s + n_p^2\g2_p + 2 n_sn_p (1 - \rho_k) \right),
\label{eq:C0}
\end{eqnarray}
where $\eta_{12}:=\eta_1\eta_2$, $N_{i,m} := a_{i,m}^\dagger a_{i,m}$, $N_i := \sum_m N_{i,m}$, and  $\rho_k := \expect{N_{1,k}}/\expect{N_1}$. 
We also used 
$\no{ N_1 N_2 } = \expect{N_1}\expect{N_2}$, 
$\expect{ N_1 } = n_s$, $\expect{ N_2 } = \expect{N_{2,k}} = n_p$, 
$\no{ N_1^2 } = n_s^2 \g2_s$, and $\no{ N_2^2 } = n_p^2 \g2_p$, 
where $n_{s(p)}$ and $\g2_{s(p)}$ are an average photon number and intensity correlation function of the target~(probe) light. 
Here, $\rho_k$ is the probability of the $k$-portion of the target photon that is to be estimated.
Since $P_{\infty}$ is equal to the first term in Eq.~(\ref{eq:N}),
$\Delta_k$ is given by 
\begin{equation}
    \Delta_k = 
    \frac{\eta_{12}}{2} \expect{ N_{1,k}}\expect{N_{2,k}} 
    = \frac{\eta_{12} n_s n_p}{2} \rho_k.  
        \label{eq:depth3}
\end{equation}
As the coefficient of $\rho_k$ is independent of $k$, 
the target state $\hat{\rho}$ can be reconstructed from the
HOM experiments with various settings of $k$. 
Furthermore, we note that Eq.~(\ref{eq:depth3}) holds even when dark counts of the detectors are considered, 
assuming that the contribution of the dark counts is the same between $P_0$ and $P_\infty$.
When the target light is a polarization qubit, $\rho_k$ corresponds to the probability obtained from projective measurements, as follows: 
\begin{equation}
 \rho_k = \bra{k}\hat{\rho}\ket{k}, 
\end{equation}
where $\ket{k}$ is the $k$-polarized single-photon state. By measuring $\Delta_k$ for four independent polarization configurations, namely horizontal~($H$), vertical~($V$), diagonal~($D$), and right-circular~($R$) polarizations, we can reconstruct the target state. The feature of HOM-QST is that no prior knowledge of the probe light is needed, such as $n_p$ and $\g2_{p}$ in $P_\infty$ and $P_k$; moreover, there is no theoretical requirement on these values.

Next, we consider a more practical situation wherein there exists a non-zero mode mismatch between the target and probe photons even when both are in state $k$. 
We describe the operators for the target and probe photons that are mode-matched to the target by the same notations as those used above. 
Moreover, we introduce operators for the probe light that express perfectly distinguishable mode mismatches with the target photon 
by adding overlines to the related operators. 
Eq.~(\ref{eq:povm}) is expanded to include the mode mismatch as 
\begin{equation}
P = \eta_{12}\no{  \sum_m (N_{1,m}^\U{out}+\overline{N}_{1,m}^\U{out})  \sum_{m'}( N_{2,m'}^\U{out}+\overline{N}_{2,m'}^\U{out}) }. \nonumber
\label{eq:povm2}
\end{equation}
In input modes, 
$\expect{ N_1 } = n_s$ and $\expect{\overline{N}_1}=0$ hold for the target photon, and $\expect{ N_{2,k} + \overline{N}_{2,k} } = n_p$ and 
$\no{ (N_{2,k} + \overline{N}_{2,k})^2 } = n_p^2\g2_p$ hold for the probe light. 
By defining the mode overlap probability between the target and probe light photons as $M := \expect{N_{2,k}}/n_p$, 
the coincidence probability $P_k$ 
is described as 
\begin{eqnarray}
P_k&=& \frac{\eta_{12}}{4}\left(
\no{ (N_1 + N_{2,k} + \overline{N}_{2,k})^2 } 
- 2 \expect{ N_{1,k}}\expect{N_{2,k}}
\right) \nonumber\\
\label{eq:Ck_MM} \\
&=& \frac{\eta_{12}}{4}\left(
n_s^2\g2_s +n_p^2\g2_p + 2 n_sn_p (1 - M \rho_k)\right). 
\label{eq:C'}
\end{eqnarray}
Note that the case of $M = 1$ in Eq.~(\ref{eq:C'}) corresponds with $P_k$ in Eq.~(\ref{eq:C0}), and the case of $M = 0$ corresponds with
\begin{eqnarray}
    P_\infty &=&
    \frac{\eta_{12}}{4}\no{ (N_1+N_{2,k}+
    \overline{N}_{2,k})^2} \label{eq:Pinfty} \\
    &=& \frac{\eta_{12}}{4}\left(n_s^2\g2_s + n_p^2\g2_p + 2 n_sn_p \right). 
    \label{eq:Cinf}
    \end{eqnarray}
The depth $\Delta_k$ of the HOM dip is described as 
\begin{eqnarray}
\Delta_k = \frac{\eta_{12} n_s n_p M}{2} \rho_k. 
\label{eq:Ak}
\end{eqnarray}
Since $M$ is independent of $k$, we can reconstruct $\hat{\rho}$ without estimating $M$.

\begin{figure*}[t]
 \begin{center}
\scalebox{1}{\includegraphics{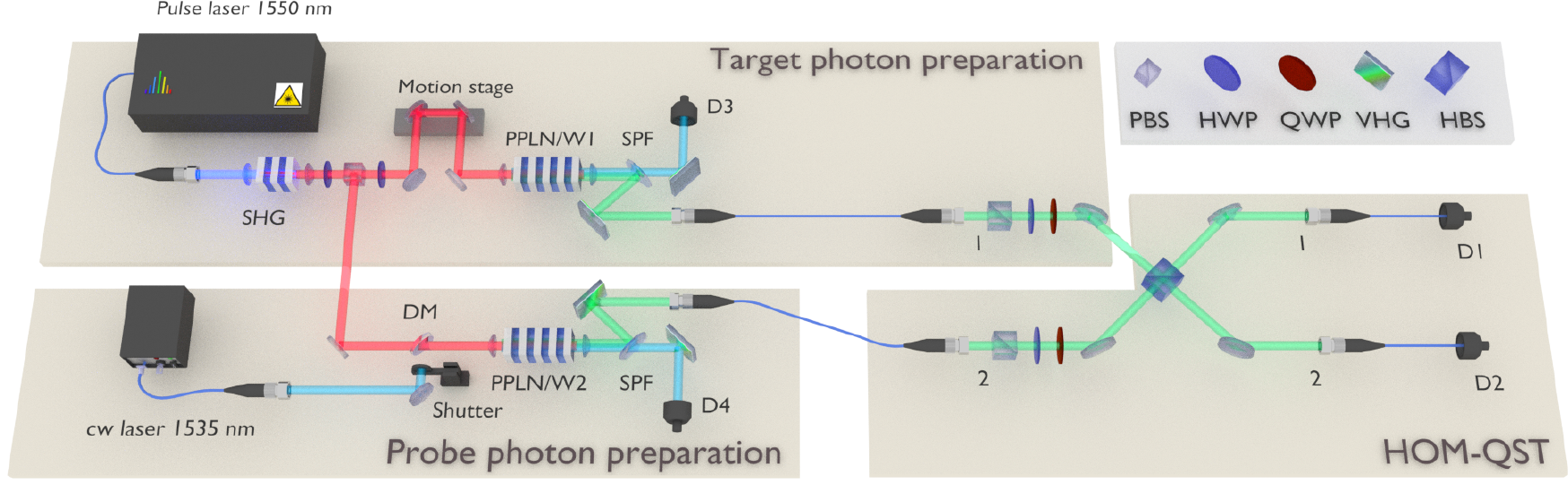}}
  \caption{The experimental setup. The setup consists of two stages, namely preparation of the target photon and probe light as well as preparation of the HOM interferometer for HOM-QST. In the first stage, the HSPS is generated 
as the target photon by SPDC. The probe light is prepared in a HSPS or a thermal state via SPDC with or without using heralding signals from D4, or in a coherent state via difference frequency generation. SPF: short-pass filter; DM: dichroic mirror; SHG: second harmonic generation; PPLN/W; periodically poled lithium niobate waveguide; VHG: volume holographic grating. 
  \label{fig:Experiment}} 
 \end{center}
\end{figure*}
Finally, we consider the BS imbalance and state dependency of the detection efficiency. 
The transformations 
of the BS with transmittance $T$ and reflectance $R$
are as follows: $N_{1,m}^\U{out}\rightarrow (\sqrt{T}a_{1,m}^\dagger + \sqrt{R}a_{2,m}^\dagger) (\sqrt{T}a_{1,m} + \sqrt{R}a_{2,m})$ 
and 
$N_{2,m'}^\U{out}\rightarrow (\sqrt{R}a_{1,m'}^\dagger - \sqrt{T} a_{2,m'}^\dagger) (\sqrt{R} a_{1,m'} - \sqrt{T}a_{2,m'})$. 
In addition, to treat the state-dependent detection efficiencies, 
$\eta_{1}$, $\eta_{2}$, and $\eta_{12}=\eta_1\eta_2$ in Eq.~(\ref{eq:povm}) are replaced with 
$\eta_{1m}$, $\eta_{2m'}$, and $\eta_{mm'}=\eta_{1m}\eta_{2m'}$, respectively.  
A straightforward calculation is thus derived as 
\begin{align}
P_k&=\sum_{m,m'}\eta_{mm'}\left\langle: (T N_{1,m} + R N_{2,m} + R \overline{N}_{2,m}) \right. \nonumber \\
&\times \left.(R N_{1,m'} + T N_{2,m'} + T \overline{N}_{2,m'}):\right\rangle 
- 2 \eta_{kk} T R n_s n_p M \rho_k. \nonumber
\end{align}
Then, the depth of the HOM dip is described as 
\begin{eqnarray}
\Delta_k = 2\eta_{kk} T R n_s n_p M \rho_k, 
\label{eq:Aetakk}
\end{eqnarray}
which shows that we can reconstruct $\hat{\rho}$ regardless of $T$ and $R$ using only the additional estimation of the relative values of the relative detection efficiencies.

In conventional QST, an $n$-partite qudit state can be estimated by preparing $n$ sets of the measurement configuration for the 1-qudit state. Similarly, $n$ sets of HOM interferometers, each with the probe lights, enable performing HOM-QST of the $n$-partite qudit state, including the entangled states. Accordingly, the extension of Eq.~(\ref{eq:Aetakk}) to the $n$-partite qudit states is described in Supplemental material. We observe that, compared with the conventional QST settings, the number of measurements in HOM-QST is the same as scaling by $d^{2n}$, while the coincidence rate changes from $\mathcal{O}(\eta^n )$ to $\mathcal{O}(\eta^{2n} n_p^n)$. 
To shorten the measurement time, larger values of $n_p$ and $\eta$ are effective as long as $n_p \eta \ll 1$ holds as a prerequisite for HOM-QST.
In addition, the use of an $n$-partitioned thermal state 
as the probe is effective. For the thermal state, the $n$-th order cross-correlation function satisfies the condition $g^{(n)}_{p}=n!$~\cite{barnett2002methods}, which implies that 
the $n$-fold coincidence probability is $n!$ times larger than that with the coherent-state probe~($g^{(n)}_{p}=1$).

\begin{table*}[t]
\begin{center}
  \begin{tabular}{|c||c|c|c|c|c|c|} \hline
Target state & $H$ & $V$ & $D$ & $A$ & $R$ & $L$  \\ \hline \hline
  Conventional method & 0.993(00)  & 0.992(00) & 0.983(00) & 0.999(00) &  0.998(00)  & 0.995(00)\\ \hline
  HOM-based QST (a) & 0.950(20) & 0.970(16) & 0.999(06) & 0.994(08) &  0.942(19)  & 0.957(17)\\ \hline  
  HOM-based QST (b) & 0.990(03) & 0.959(14) & 0.995(03) & 0.999(01) &  0.967(06)  & 0.976(05)\\ \hline
  HOM-based QST (c) & 0.996(01) & 0.969(02) & 0.996(02) & 0.994(02) &  0.975(03)  & 0.980(02)\\ \hline 
    \end{tabular}
    \caption{The fidelities estimated by the conventional 1-qubit tomography~\cite{James2001} and the HOM-based QST using (a)~the HSPSs, (b)~thermal states and (c)~coherent states.
    The digits in parentheses represent the standard deviation.}
\label{table:fidelity}
\end{center}
\end{table*}

\begin{figure*}[t]
 \begin{center}
\scalebox{0.16}{\includegraphics{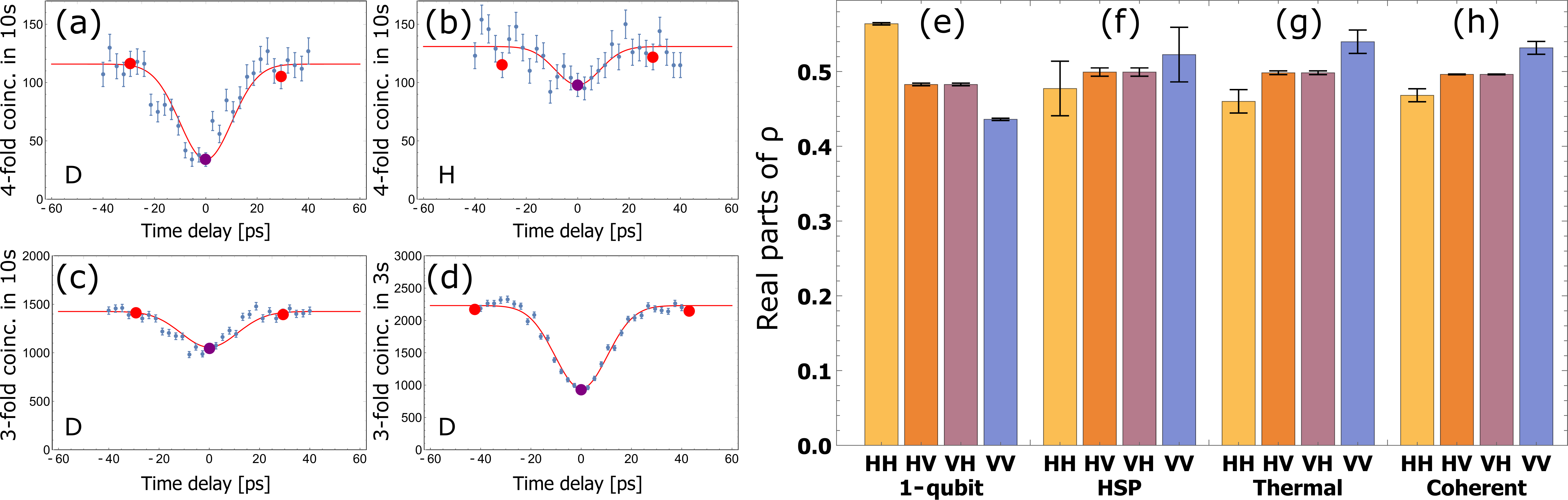}}
  \caption{(a)--(d)~Observed HOM dips between the $D$-polarized target photons and (a)~$D$-polarized HSPS, (b)~$H$-polarized HSPS, (c)~$D$-polarized thermal-state, and (d)~$D$-polarized coherent-state probes. In the HOM-QST experiment, we measure only three data points rather than observing the HOM dip for each polarization setting to estimate the coincidence count rates $C_{k}$~(purple point) at the zero delay point and $C_{k\infty}$~(two red points) at large delay points. The depth of the HOM dip is calculated as $\tilde{\Delta}_k=C_{k\infty}-C_k$. The measurement times for the data points in HOM-QST are \SI{30}{s} each for (a), (b) HSPS and (c) thermal-state probes and \SI{10}{s} for the (d) coherent-state probe. (e)--(h)~ Real parts of the reconstructed density matrices of the $D$-polarized target states using (e)~conventional 1-qubit QST and HOM-QST with (f)~HSPS, (g)~thermal-state, and (h)~coherent-state probes. Each value on the vertical axis corresponds to the real part of the matrix element, e.g., the value for label HV corresponds to $\mathrm{Re}\bra{H}\hat{\rho}\ket{V}$. The error bars indicate the standard deviations estimated using the Monte Carlo method under the assumption of Poisson statistics for the photon counts.
    \label{fig:HOMtomo}} 
 \end{center}
\end{figure*}
{\it Experiment.-}
We experimentally demonstrate HOM-QST for polarization qubits. We prepared three types of probe lights, namely the heralded-single-photon-state~(HSPS)~($\g2_p\ll 1$), thermal-state~($\g2_p\sim 2$), and coherent-state~($\g2_p\sim 1$) probes. 
The sketch of the experimental setup is shown in Fig.\ref{fig:Experiment}.  A fiber-based mode-locked laser emitting \SI{1550}{nm} optical pulses at a repetition rate of \SI{250}{MHz} is used as a pump laser.
This setup consists of an electro-optical comb generator~\cite{Wakui:20, Tsujimoto2021} followed by a high-speed pulse picker 
and a fiber amplifier. After the output is passed to the free-space system, it is frequency-doubled via second harmonic generation~(SHG) using a type-0 \SI{10}{mm}-long periodically poled lithium niobate waveguide~(PPLN/W).
Then, the SHG light at \SI{775}{nm} is divided into two spatial modes using 
a half waveplate~(HWP) and polarization beam splitter~(PBS) before being used to pump the subsequent PPLN/Ws for spontaneous parametric down-conversion~(SPDC). 
One pump beam is used to prepare the target photon; this beam is coupled with a \SI{34}{mm} long PPLN/W~(PPLN/W~1) 
and generates a \SI{1535}{nm} and \SI{1565}{nm} photon pair via SPDC. 
The \SI{1535}{nm} photon is detected by $\mathrm{D}_3$
after frequency-filtering using a volume holographic grating~(VHG) of bandwidth \SI{58}{GHz} to herald the \SI{1565}{nm} photon. 
The HSP at \SI{1565}{nm} is frequency-filtered using another VHG of bandwidth \SI{32}{GHz}, 
which is narrower than the pump bandwidth of \SI{75}{GHz}, 
and is used as the target photon. 
The other pump at \SI{775}{nm} is coupled with a second \SI{34}{mm}-long PPLN/W~(PPLN/W~2) 
and used to prepare the 1565 nm probe light. 
The HSPS probe is prepared in a similar manner to that of the target photon; 
the thermal-state probe is prepared by discarding the heralding signal in the  configuration, 
and the coherent-state probe is prepared by difference frequency generation 
of the \SI{775}{nm} light 
with the input of another \SI{1535}{nm} continuous-wave~(cw) laser beam as the pump light. 
In the second stage of the HOM-QST, the target photon and probe light are obtained from modes~1 and 2, respectively. Each photon is then initialized to the $H, V, D, A, R$, and $L$ polarization states using a HWP and quarter waveplate~(QWP) after a PBS, where  
$A$ and $L$ are the anti-diagonal and left-circular polarizations, respectively. HOM-QST is then performed by mixing the target photon and probe light
at the HBS, followed by the photon detections at $\mathrm{D}_1$ and $\mathrm{D}_2$. All the detectors are superconducting nanowire single-photon detectors~(SSPDs)~\cite{Miki:17}, whose relative joint detection efficiencies, dependent on polarization, are $\eta_{kk}/\eta_{HH}=$ 1, 1.39, 1.19, 0.77, 1.19, and 1.19 for $k=H, V, D, A, R$, and $L$, respectively. The electrical signals from the SSPDs are fed to a time-to-digital converter~(not shown). The signal from $\mathrm{D}_3$ is used as the start signal, and the signals from $\mathrm{D}_4$, $\mathrm{D}_1$, and $\mathrm{D}_2$ are used as the stop signals. For the HSPS probe, the four-fold coincidence events are used as successful events. For the thermal-state and coherent-state probes, the three-fold coincidence events among $\mathrm{D}_3$, $\mathrm{D}_1$, and $\mathrm{D}_2$ are used. 

 As a preliminary experiment, we characterized the quantum state of the target photon by the conventional 1-qubit QST~\cite{James2001}. The QST was performed by inserting a QWP, a HWP, and a PBS between the HBS and $\mathrm{D}_1$ in Fig.~\ref{fig:Experiment}, while blocking the probe light. Using the diluted maximum-likelihood algorithm~\cite{Rehacek2007}, 
we reconstructed the density operator $\hat{\rho}$ of the target photon. 
We show the real part of the reconstructed density matrix of 
the $D$-polarized target photon in Fig.~\ref{fig:HOMtomo}(e). 
We next evaluated the fidelity $F_k:=\bra{k}\hat{\rho}\ket{k}$ 
of the experimentally prepared $k$-polarized target state $\hat{\rho}$
to $\ket{k}$ for $k\in\{H,V,D,A,R,L\}$. 
These results are summarized in Table.~\ref{table:fidelity}. 
The fidelity in each case is close to unity, thereby indicating that the target states are almost ideal states. 

We show the experimental results of HOM-QST using the HSPS probe. 
For example, we show the HOM dip between the $D$-polarized target photon and $D$($H$)-polarized probe 
photon in Fig.~\ref{fig:HOMtomo}(a)((b)). 
The depth of each dip was proportional to the probability $\Delta_{D(H)}$ in Eq.~(\ref{eq:Aetakk}). 
We observed the largest count rate as $\tilde{\Delta}_D = \SI{6.1}{Hz}$ and the smallest count rate as $\tilde{\Delta}_A=\SI{-0.7}{Hz}$, 
indicating that the target state is close to $\ket{D}$. The count rates for the other probe states are provided in the Supplemental material.
The negative value of the depth, such as $\tilde{\Delta}_A$, occurs 
when the HOM dip is almost zero because of statistical fluctuations of the photon counts and optical misalignment. 
The latter effect could be removed
by calibrating the dependency of the photon count rate on the optical delay. In the demonstration, we reconstructed the density operator based on two strategies to treat the negative value as either zero or using only the results of the positive values to avoid the negative values. 
For each case, 
the fidelity was estimated as $F_D=0.999(06)$,  
and the difference between the two strategies was negligibly small. 
Below, we show the results based on the latter strategy. 
The real part of the reconstructed density matrix of the $D$-polarized target photons is shown in Fig.~\ref{fig:HOMtomo}(f). 
The fidelities are summarized in Table~\ref{table:fidelity}. 
Each fidelity value is close to that estimated by 1-qubit QST, 
thus clearly showing that HOM-QST provides reliable estimations of 
the input quantum states. 
We show the reconstructed density matrices for the other target states in the Supplemental material.

Finally, we show the experimental results of HOM-QST 
using the thermal-state and coherent-state probes. 
The HOM dip with the $D$-polarized target photon 
and $D$-polarized thermal~(coherent)-state probe is 
shown in Fig.~\ref{fig:HOMtomo}(c)((d)). 
These HOM visibilities are smaller than that in the case of the HSPS probe 
in Fig.~\ref{fig:HOMtomo}(a)
owing to the difference between $n_p$ and $\g2_p$~(see Supplemental material). 
Nonetheless, 
the reconstructed density operators and their fidelities 
are similar for all cases, as shown in Table~\ref{table:fidelity}
and Figs.~\ref{fig:HOMtomo}(g) and (f). 
These results elucidate our theoretical prediction that precise estimations can be achieved by HOM interference 
regardless of the values of $n_p$ and $\g2_p$.

{\it Conclusion.-} In conclusion, 
we proposed the QST of photonic qudits via HOM interference and demonstrated it successfully using polarization qubits as the proof-of-principle.
The proposed method shifts the complexity of the measurement system 
to the probe light preparation system, which 
then enables us to construct the optical circuit with only passive elements.
We theoretically showed that 
the HOM-QST does not need information regarding the probe light, 
such as intensity and photon statistics, 
and allows for experimental imperfections, 
such as mode mismatch between the photons 
and use of an unbalanced BS. 
We demonstrated the HOM-QST using the HSPS, thermal-state, and coherent-state probes. The estimated fidelities were of a similar order as those obtained 
via conventional QST, which demonstrates the 
precise estimation of the density operator regardless of 
the intensity and photon statistics of the probe light. 
In the demonstration, we applied HOM-QST to the polarization qubits. 
However, our method exhibits its power for high-dimensional systems in which the measurement setup tends to be complex.  
For example, in the case of frequency qudits, a line-by-line waveform shaping technique for frequency combs~\cite{Cundiff2010} can be used to prepare the probe light without requiring any nonlinear optical interactions in a measurement setup, e.g. as shown in Refs.~\cite{Kobayashi2016,Gil-Lopez:21,Kues2019}. 
Our method serves as a new alternative to perform QST with a passive measurement setup and is expected to be applicable to integrated quantum photonics.

Y.T. is supported by JSPS KAKENHI Grant No. JP20K14393. 
Y.T. and K.W. are supported by CREST, JST JPMJCR1772.
R.I. and T.K. are supported by CREST, JST JPMJCR1671, and JSPS KAKENHI Grant Nos. JP21H04445 and JP20H01839.

\bibliographystyle{h-physrev}

\newpage 
\mbox{ }
\newpage
\begin{widetext}

\renewcommand{\thefigure}{S\arabic{figure}}
\setcounter{table}{0}
\renewcommand{\thetable}{S\arabic{table}}
\setcounter{figure}{0}
\renewcommand{\theequation}{S\arabic{equation}}
\setcounter{equation}{0}

\section{Supplemental material}

\section{HOM-QST of a composite system}
We consider the HOM-QST of an $n$-qudit system. For convenience, we treat the detection efficiencies of the photon detectors as constants in the derivation and
consider their state dependencies in the end.
The setup for $n$-qudit tomography consists of $n$ sets 
of HOM interferometers. 
We label the input ports of the target and probe light for the $i$-th interferometer as $2i-1$ and $2i$, respectively. 
We define 
the set of serial numbers of the HOM interferometers as $X_n := \{ 1,\ldots, n\}$. 
We also define the power set of $X_n$ as ${\mathfrak P}(X_n)$ 
to separate the interferometers into two groups: interferometers in $S\in {\mathfrak P}(X_n)$ satisfy the condition of zero time delay between the target and probe light, and those in $X_n\backslash S$ satisfy the condition of large time delays. 
We denote the list of probe states affecting the coincidence probabilities as $\kappa(S) := k_{1},\ldots, k_{n}$, 
where 
$k_i$ is the probe state of the $i$-th interferometer for $i\in S$, 
and $k_i = \emptyset$ for $i\in X_n\backslash S$. 
Based on these definitions, the 2$n$-fold coincidence probability $P_{\kappa(S)}^{(n)}$ 
is given by extending Eq.~(\ref{eq:Ck_MM}) in the main text as 

\begin{align}
P_{\kappa(S)}^{(n)}
&= \frac{\eta}{4^n}
\left\langle:
\left( \prod_{i\in X_n \backslash S} (N_{2i-1}+N_{2i,k_{i}}+\overline{N}_{2i,k_{i}})^2\right)\left( \prod_{j\in S} (N_{2j-1}+N_{2j,k_{j}}+\overline{N}_{2j,k_{j}})^2 
- 2 N_{2j-1,k_{j}}N_{2j,k_{j}} \right) : \right\rangle \nonumber \\
&=
\sum_{S'\in {\mathfrak P}(S)} 
(-1)^{|S'|} \Delta_{\kappa(S')}^{(n)}, 
\label{eq:n} 
\end{align}

where $\eta$ is the total detection efficiency from $2n$ photon detectors, and 

\begin{align}
\Delta_{\kappa(S')}^{(n)} &:=  
\frac{\eta}{4^n}
\left\langle:\left( \prod_{i\in X_n \backslash S'} (N_{2i-1}+N_{2i,k_{i}}+\overline{N}_{2i,k_{i}})^2 \right)\left( 2^{|S'|} \prod_{j\in S'}N_{2j-1,k_{j}}N_{2j,k_{j}} \right) :\right\rangle
, 
\label{eq:deltan}
\end{align}

with $\prod_{i(j)\in \varnothing} \cdots := 1$. 
$P_{\kappa(\varnothing)}^{(n)} = \Delta_{\kappa(\varnothing)}^{(n)}$ is 
the probability obtained when the photons in all the interferometers are distinguishable corresponding to $P_\infty$ in Eq.~(\ref{eq:Pinfty}) in the main text for $n=1$. 
We note that 
$\Delta_{\kappa(X_n)}^{(n)}$, included in 
the observable probability $P_{\kappa(X_n)}^{(n)}$, 
contains information pertaining to

\begin{eqnarray}
\rho_{\kappa(X_n)} &:=& \expect{ \prod_{j\in X_n}N_{2j-1,k_{j}} }/n_{s,n} \\
&\sim & \bra{k_1\cdots k_n}\hat{\rho}\ket{k_1\cdots k_n}, 
\end{eqnarray}
which is to be determined,  
where $n_{s,n}:=\expect{ \prod_{j\in X_n}N_{2j-1}}$ 
is the $n$-th order cross-correlation of the target state. 
$\Delta_{\kappa(X_n)}^{(n)}$ is given by 
\begin{eqnarray}
\Delta_{\kappa(X_n)}^{(n)}
&=& 
\frac{\eta}{4^n}
\no{
\left( 2^{|X_n|} \prod_{j\in X_n}N_{2j-1,k_{j}}N_{2j,k_{j}} \right) 
}  \\
&=& 
\frac{\eta}{2^{n}}
\expect{
\prod_{j\in X_n}N_{2j-1,k_{j}} }
\expect{
\prod_{j\in X_n}N_{2j,k_{j}} }.
 \label{eq:DeltaXn1}
\end{eqnarray}
Similar to the expansion of Eq.~(\ref{eq:depth3}) to Eq.~(\ref{eq:Aetakk}) in the main text for $n=1$, 
by considering the mode mismatch effects, state dependency of the detection efficiency, and 
BS imbalance, Eq.~(\ref{eq:DeltaXn1}) can be expressed as 
\begin{eqnarray}
 \Delta_{\kappa(X_n)}^{(n)}=2^nT^nR^n\eta_{X_n} n_{s,n} n_{p,n} M_{X_n} \rho_{\kappa(X_n)}, 
\label{eq:DeltaXn}
\end{eqnarray}
where $n_{p,n}:=\expect{\prod_{j\in X_n} (N_{2j,k_{j}}+\overline{N}_{2j,k_{j}})}
= g_{p,X_n}^{(n)}\prod_{j\in X_n} \expect{N_{2j,k_{i}}+\overline{N}_{2j,k_{j}}}$ 
is the $n$-th order cross-correlation of the probe state, 
$\eta_{X_n}:=\prod_{j\in X_n}\eta_{k_j}\eta_{k_j}$ is 
the total state-dependent detection efficiency of the $2n$ photon detectors,
$M_{X_n}:=\prod_{j\in X_n} M_j$ is the product of mode-matching 
between the target and probe light in the $j$-th interferometer with 
$M_j:=\expect{N_{2j,k_{j}}}/\expect{N_{2j,k_{j}}+\overline{N}_{2j,k_{j}}}$, 
and $g_{p,X_n}^{(n)}=
n_{p,n}/\prod_{j\in X_n} \expect{N_{2j,k_j}+\overline{N}_{2j,k_j}}
= \expect{ \prod_{j\in X_n} N_{2j,k_j} }/\prod_{j\in X_n} \expect{N_{2j,k_j} }$ 
is the $n$-th order cross-correlation function of the probe light. 
The coefficient of $\rho_{\kappa(X_n)}$ in the equation does not depend on $\kappa(X_n)$. 
Thus, we can reconstruct $\hat{\rho}$ from $(1+d^2)^n$ different experimental settings. 

We present examples for the $n=1$ and $n=2$ cases, as follows. 
From Eqs.~(\ref{eq:n}) and (\ref{eq:deltan}), 
for $n=1$, we see that 
$P_{\kappa(X_1)}^{(1)}=P_{k_1}^{(1)} = \Delta_{\kappa(\varnothing)}^{(1)} - \Delta_{k_1}^{(1)}$, 
where $\Delta_{\kappa(\varnothing)}^{(1)} = \Delta_{\emptyset}^{(1)} = P_\infty$ in Eq.~(\ref{eq:Pinfty}) and $\Delta_{k_1}^{(1)} = \eta n_{s,1}n_{p,1}M_1\rho_{k_1}/2$ is equivalent to Eq.~(\ref{eq:Ak}). 
For $n=2$, the four-fold coincidence probability $P_{\kappa(X_2)}^{(2)} = P_{k_1,k_2}^{(2)}$ can be expressed as
\begin{eqnarray}
P_{k_1,k_2}^{(2)}
&=&\Delta_{\kappa(\varnothing )}^{(2)} - \Delta_{\kappa(\{ 1\})}^{(2)} - \Delta_{\kappa(\{ 2\})}^{(2)} 
+ \Delta_{\kappa(\{ 1,2\})}^{(2)} \\
&=&
\Delta_{\emptyset,\emptyset}^{(2)}
+ (P_{k_1,\emptyset}^{(2)} - \Delta_{\emptyset,\emptyset}^{(2)})
+ (P_{\emptyset,k_2}^{(2)} - \Delta_{\emptyset,\emptyset}^{(2)})
+ \Delta_{k_1,k_2}^{(2)} \\
&=&
P_{k_1,\emptyset}^{(2)} + P_{\emptyset, k_2}^{(2)} 
- \Delta_{\emptyset,\emptyset}^{(2)} + 4 n_{s,2} n_{p,2}M_1M_2
\rho_{k_1,k_2}.
\label{eq:2qubits}
\end{eqnarray}
Thus, $\rho_{k_1,k_2}$ is estimated by measuring 
$P_{k_1,k_2}^{(2)}$, $\Delta_{\emptyset,\emptyset}^{(2)}$, $P_{k_1,\emptyset}^{(2)}$ and $P_{\emptyset,k_2}^{(2)}$. 

\section{HOM visibility}
We consider the HOM visibility between the $k$-polarized target photon and probe light. From Eqs.~(\ref{eq:Cinf}) and (\ref{eq:Ak}) in the main text, the HOM visibility is expressed as 
\begin{equation}
V\u{th}:=\frac{\Delta_k}{P_\infty}=M\frac{1}{1+\frac{\zeta\g2_p+1/\zeta\g2_s}{2}}, 
\label{eq:HOMvis}
\end{equation}
where $\zeta := n_p/n_s$. 
In Figs.~3(a), (c) and (d), 
we show the HOM dips between the $D$-polarized target photons and $D$-polarized probes. The experimental HOM visibilities defined by 
$V\u{ex}:= (C_{k\infty}-C_k)/C_{k\infty}$ are given by $V\u{ex}=0.707(57)$, $0.311(29)$, and $0.575(16)$ for the HSPS, thermal-state, and coherent-state probes, respectively. 
In the HOM experiment with the HSPS probe, we additionally measured the experimental parameters as 
$\zeta=0.694(01)$, $\g2_s=0.211(02)$, and $\g2_p=0.355(03)$. 
By substituting these values into Eq.~(\ref{eq:HOMvis}), 
and assuming $V\u{th}=V\u{ex}=0.707(57)$,  $M=0.901(73)$ was estimated. For the thermal-state probe, 
$\zeta=0.0646(06)$ was observed. Using the estimated value of $M$ and assuming that $\g2_p=2$, we obtained $V\u{th}=0.334(27)$. 
In the coherent-state probe experiment, $\zeta=0.382(01)$ and $\g2_s=0.191(03)$ were observed. Assuming $\g2_p=1$, we obtained $V\u{th}=0.626(51)$. These values are in good agreement with $V\u{ex}$ within the margin of errors, which indicates that our model successfully explains the influence of $\zeta$ and $g^{(2)}_{s(p)}$ 
on HOM visibility. 

\section{Count rates}
We list all the count rate data for the conventional QST and HOM-QST in Table~\ref{table:HOMcounts}. 
\begin{table*}[h]
\begin{center}
  \begin{tabular}{|c|c||c|c|c|c|c|c|} \hline
& & \multicolumn{6}{c|}{Measurement polarization $k$}\\ \cline{3-8}
QST method & Probe & $H$ & $V$ & $D$ & $A$ & $R$ & $L$  \\ \hline \hline
Conventional QST & - & 10885 & 8417  & 18969 & - &  9910  & -\\ \hline
& HSP & 2.3 & 2.5  & 6.1 & -0.7 &  2.7  & 1.9\\ \cline{2-8}
HOM-QST & Thermal state & 11.5 & 14.6 & 32.2 & -1.9 &  13.1  &  13.8\\ \cline{2-8}
 & Coherent state & 145.3 & 166.3 & 391.7 & -84.8 &  197.7& 129.1\\ \hline
   \end{tabular}
    \caption{ Observed count rates per \SI{1}{s} for the $D$-polarized target photon.
    For conventional QST, single count rates are listed. 
    For HOM-QST, $\tilde{\Delta}_k/\eta_{kk}$ is listed. 
    The measurement polarization $k$ is chosen using the WPs and PBS for conventional QST and probe light polarization for HOM-QST. 
    }
\label{table:HOMcounts}
\end{center}
\end{table*}

\section{Results of HOM-QST}
We present all the density matrices reconstructed using 1-qubit tomography and HOM-QST in Fig.~\ref{fig:sup_density}. 
\begin{figure*}[t]
 \begin{center}
\scalebox{0.36}{\includegraphics{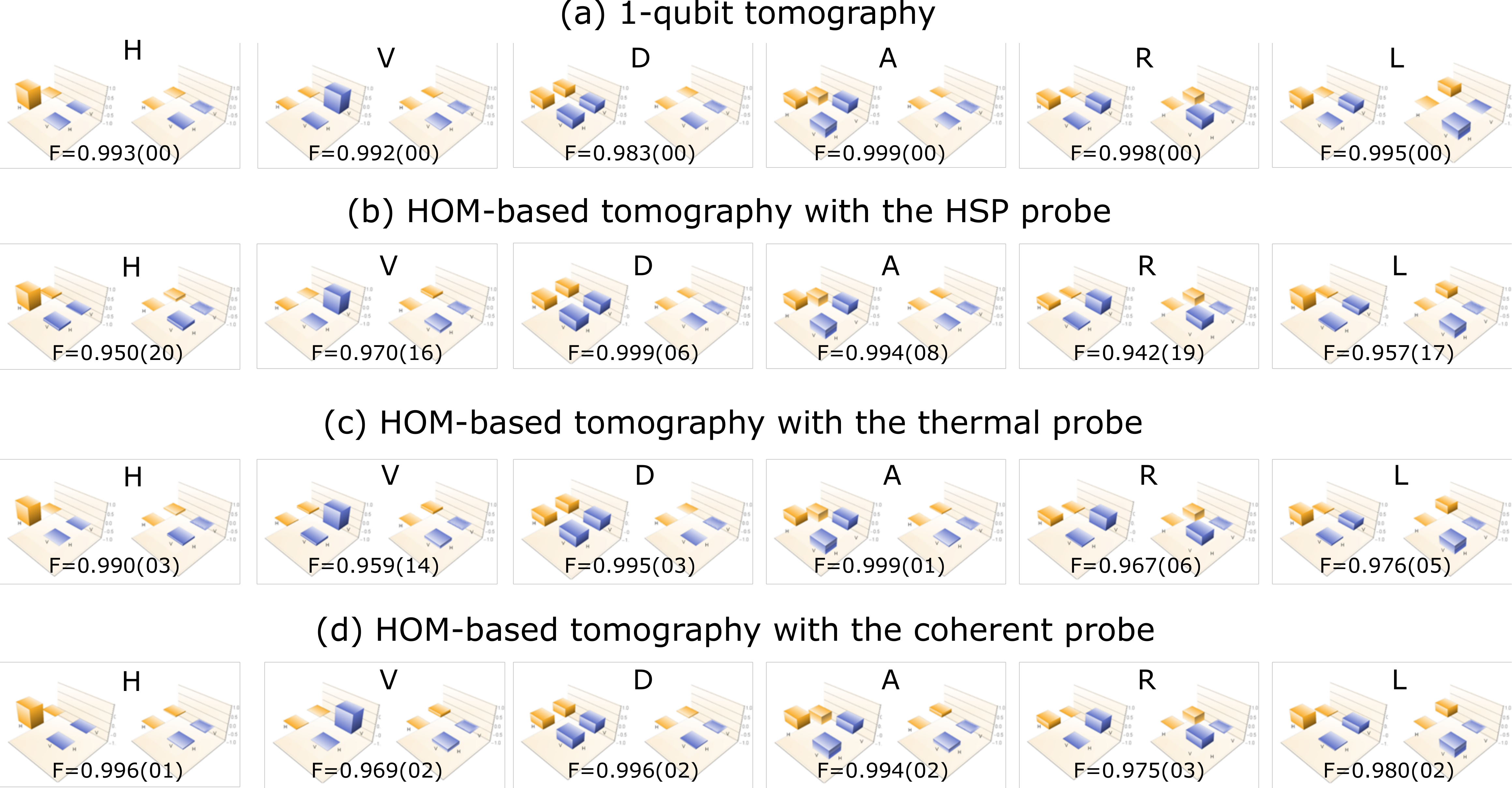}}
  \caption{Density matrices reconstructed using (a)~ 1-qubit tomography as well as HOM-QST with (b) HSPS, (c) thermal-state, and (d) coherent-state probes. The left and right parts of each figure correspond to the real and imaginary parts of the reconstructed density matrix, respectively. 
    \label{fig:sup_density}} 
 \end{center}
\end{figure*}
\end{widetext}

\end{document}